\documentclass[lettersize,journal]{IEEEtran}
\usepackage{amsmath,amsfonts}
\usepackage{algorithmic}
\usepackage{algorithm}
\usepackage{array}
\usepackage[caption=false,font=normalsize,labelfont=sf,textfont=sf]{subfig}
\usepackage{textcomp}
\usepackage{stfloats}
\usepackage{url}
\usepackage{verbatim}
\usepackage{graphicx}
\usepackage{cite}
\usepackage{xcolor}
\begin{document}

\title{Balancing Exploration and Cybersickness: Investigating Curiosity-Driven Behavior in Virtual Environments}

\author{Tangyao Li, and Yuyang Wang,~\IEEEmembership{Member,~IEEE}
        % <-this % stops a space
\thanks{ Tangyao Li (email: tli724@connect.hkust-gz.edu.cn) and Yuyang Wang (email: yuyangwang@hkust-gz.edu.cn) are with the Information Hub, the Hong Kong University of Science and Technology (Guangzhou), China, 511453.}% <-this % stops a space
\thanks{Corresponding author: Yuyang Wang}
%\thanks{E-mail: tli724@connect.hkust-gz.edu.cn, yuyangwang@hkust-gz.edu.cn}
}

\markboth{Journal of \LaTeX\ Class Files,~Vol.~14, No.~8, August~2021}%
{Shell \MakeLowercase{\textit{et al.}}: A Sample Article Using IEEEtran.cls for IEEE Journals}


\maketitle

\begin{abstract}

Virtual reality offers the opportunity for immersive exploration, yet it is often undermined by cybersickness. However, how individuals strike a balance between exploration and discomfort remains unclear. Existing method (e.g., reinforcement learning) often fail to fully capture the complexities of navigation and decision-making patterns. This study investigates how curiosity influences users' navigation behavior, particularly how users strike a balance between exploration and discomfort. We propose curiosity as a key factor driving irrational decision-making and apply the free energy principle to model the relationship between curiosity and user behavior quantitatively. Our findings indicate that users generally adpot conservative strategies when navigating. Also, curiosity levels tend to rise when the virtual environment changes.These results illustrate the dynamic interplay between exploration and discomfort. Additionally, it offers a new perspective on how curiosity drives behavior in immersive environments, providing a foundation for designing adaptive VR environments. Future research will further refine this model by incorporating additional psychological and environmental factors to improve prediction accuracy.
\end{abstract}


\section{Introduction}
Virtual reality (VR) enables immersive experiences that transcend the boundaries of time and space, with virtual navigation serving as a core component of this immersion~\cite{chardonnet2017features}. Through navigation, users explore environments and derive rewards such as completing tasks, discovering new spaces, or enjoying the immersive experience~\cite{hung2024exploring,kim2022falling}. However, these experiences are shaped by a complex interplay between curiosity and potential risks, particularly cybersickness~\cite{venkatakrishnan2024effects}. This interplay profoundly impacts user behavior, as curiosity drives exploration of novel environments, while the risk of discomfort often tempers such impulses. Understanding this balance is crucial for designing VR systems that are both engaging and comfortable.

Curiosity is a key driver of exploratory behavior in VR, often prompting users to seek out novel and uncertain experiences~\cite{arnone2011curiosity}. However, individual differences, such as prior VR experience and personal risk tolerance, lead to diverse interaction patterns~\cite{davis2014systematic,wang2021using}. While some users eagerly explore, others adopt conservative strategies to avoid risks, particularly in environments with a high likelihood of cybersickness. This highlights the importance of investigating how curiosity and reward interact to shape user decision-making. Insights into this relationship could boost the development of personalized VR applications that balance engagement with user comfort~\cite{wang2023modeling}.

Despite its significance, the dynamic relationship between curiosity and cybersickness remains underexplored. Cybersickness arises from sensory conflicts between visual input and bodily perception during spatial locomotion in VR, leading to symptoms such as dizziness, nausea, and eye strain~\cite{reason1975motion}. These symptoms are prevalent, with 80\% to 95\% of VR users experiencing some degree of discomfort, and 5\% to 30\% abandoning sessions due to severe symptoms~\cite{arns2005relationship}. While much research has focused on the prevalence and causes of cybersickness, the role of curiosity-driven exploration in exacerbating or mitigating discomfort has received limited attention~\cite{arnone2011curiosity}. Addressing this gap is essential for improving user experience in VR.

Behavior in VR is further complicated by the inherent uncertainty of virtual environments and the brain's limited capacity to process real-time information~\cite{augier2002administrative}. This often leads to irrational decisions, particularly when users face conflicts between curiosity-driven exploration and reward-seeking behavior~\cite{gomez2019shinobi}. Sensory conflicts can amplify this tension, creating a complex dynamic where curiosity, reward, and discomfort intersect. Understanding these behavioral strategies is key to addressing both the cognitive and physiological challenges users encounter in VR.

Analyzing navigation speed profiles offers a novel approach to studying these dynamics. By examining how users balance curiosity and reward during navigation, we can gain insights into the factors influencing irrational decisions and the severity of cybersickness. Notably, in the context of the current study, the concept of ``reward" is not modeled as a beneficial outcome but as a measure of the severity of cybersickness experienced during virtual navigation. This reframing highlights the tension between curiosity-driven exploration and the discomfort associated with sensory conflicts, offering a unique perspective on user behavior in VR.

Ultimately, the severity of cybersickness is closely tied to how users negotiate the trade-off between immersive rewards and discomfort~\cite{wang2021development}. Investigating this ``reward-curiosity conflict" through user navigation patterns offers critical insights into the mechanisms driving behavior in VR~\cite{gomez2019shinobi}. By modeling exploratory behavior, this research aims to advance VR interaction design, reduce cybersickness, and improve user experience. The contributions are as follows:

\begin{itemize}
    \item  We employed a machine learning framework, the inverse free energy principle (iFEP), to estimate the internal variables underlying decision-making processes and to model user behavior and interaction dynamics in VR environments.
    \item By applying the iFEP method to time-series data from VR navigation, we identified key variables such as curiosity and the modeled ``reward" (quantified as the severity of cybersickness). This approach offers novel insights into how users negotiate the trade-off between curiosity-driven exploration and the risk of discomfort.
\end{itemize}

Section \ref{sec:related work} reviews studies on individual differences in behavior during virtual navigation, identifying gaps this research addresses. Section \ref{sec:methods} outlines the methodological framework, including the application of the inverse free energy principle (iFEP). Section \ref{sec:results} presents and interprets the findings, alongside an evaluation of alternative approaches. Section \ref{sec:discussion} discusses the implications of the results for understanding VR user behavior and interaction design.

\section{Related Work}\label{sec:related work}

\subsection{User behavior in VR}

User behavior in virtual environments varies widely due to individual differences. VR provides a controlled platform for experiments, where participants’ movements can be continuously recorded, enabling the collection of validated datasets for comprehensive analysis~\cite{yaremych2019tracing}. These datasets are instrumental in identifying the factors that influence user behavior during virtual navigation, which in turn impact the user experience \cite{anwar2025metaverse}.

Proxemic information, derived from user positioning, is critical for quantifying behaviors such as social apporach and avoidance \cite{dechant2017potential}. Won et al. \cite{won2016identifying} leveraged head orientation data to analyze scanning behavior in a virtual classroom, which could indicate student anxiety levels, while Gillath et al. \cite{gillath2008can} used proxemic patterns to identify prosocial approach behaviors. In addition, Javdani Rikhtehgar et al. \cite{javdani2023personalizing} examined the use of eye gaze to understand the user's information-seeking behavior in an exhibition.

% For example, visual gaze analysis has been used to replace traditional controllers for object selection, offering valuable insights into interaction dynamics~\cite{rubo2018virtual}. Similarly, head orientation is often used to estimate gaze direction in social scenarios, and combined changes in head and hand positions can reveal distinct behavioral patterns~\cite{yaremych2019tracing}. Proxemic information, derived from user positioning, is critical for quantifying behaviors such as social approach and avoidance~\cite{dechant2017potential}. For instance, Won et al. \cite{won2016identifying} leveraged head orientation data to analyze scanning behavior in a virtual classroom, which could indicate student anxiety levels, while Gillath et al. \cite{gillath2008can} used proxemic patterns to identify prosocial approach behaviors.

Comparisons between VR and real-world proxemic patterns have revealed both similarities and differences. Bailenson et al.\cite{bailenson2001equilibrium} reported that proxemic behaviors in VR often mirror those in physical settings. Additionally, walking speeds in virtual environments generally align with real-world speeds~\cite{deb2017efficacy}, although Iryo-Asano et al.\cite{iryo2018applicability} found that maximum walking speeds tend to be lower in VR. Notably, Wei et al. \cite{wei2019effects} reported that users with greater technological experience exhibit higher immersion and confidence, leading to more effective navigation behaviors.

\subsection{The role of curiosity in decision-making}

Curiosity plays a pivotal role in human decision-making, especially in environments where uncertainty-driven exploration occurs, even in the absence of external incentives and stimuli \cite{arnone2011curiosity,schmidhuber1991curious,berlyne1966curiosity}. In VR settings, curiosity has been shown to motivate users to engage more deeply with their surroundings, fostering active exploration.

Empirical studies shed light on the neural and behavioral dynamics of curiosity. For instance, Jepma et al.\cite{jepma2012neural} demonstrated that curiosity activates brain regions associated with conflict and arousal, while resolving curiosity triggers reward-related neural pathways. Over time, the impact of curiosity on decision-making evolves, as prolonged information gaps can diminish users' anticipation and reduce their exploratory drive~\cite{noordewier2017curiosity}. To address such issue, Tu and Lee \cite{tu2025curiosity} found that interest-driven curiosity had a strong positive influence on engagement. Despite its potential to introduce discomfort in uncertain situations, individuals are naturally drawn to the unknown\cite{noordewier2017curiosity}.

To model and investigate curiosity, various methodologies have been employed \cite{gomez2019shinobi, noordewier2017curiosity}. The free energy principle (FEP), introduced by Karl Friston under the Bayesian brain hypothesis~\cite{2009The,friston2010free,friston2017active}, frames decision-making as the optimization of both reward-seeking and curiosity-driven behaviors. While FEP provides a framework for understanding these dynamics, a limitation lies in the assumption that curiosity and rewards are weighted equally, which does not fully reflect real-world behavior. For example, Konaka and Naoki \cite{konaka2023decoding} noted that traditional FEP models assume constant curiosity, neglecting the dynamic nature of user curiosity, which fluctuates over time. Similarly, Millidge et al.\cite{millidge2021whence} criticized the oversimplification of curiosity in bi-choice tasks, where the interplay between curiosity and rewards is more complex and context-dependent.

Machine learning (ML) and reinforcement learning (RL) offer an alternative approach to modeling curiosity \cite{schwartenbeck2019computational}. While RL focuses on maximizing future rewards through a series of actions, it also distinguishes between extrinsic rewards (externally set) and intrinsic rewards (self-generated by the agent) \cite{chen2025extrinsic}. Notice that the ``reward'' mentioned here refers to the terminology in RL, which is different from the ``reward'' discussed in the later section that relates to the terminology in iFEP. RL-based models of curiosity-driven exploration are typically divided into novelty-based and prediction-error algorithms. In novelty-based models, curiosity motivates the brain to explore new, unfamiliar tasks, while prediction-error models suggest that curiosity is driven by the need to improve predictions about the environment~\cite{schmidhuber1991curious}. Despite the usefulness of these models, they fall short of capturing the complexity of user curiosity in immersive VR settings, where exploration is often intentional and guided by the user’s curiosity, not a passive, random process as often modeled in RL~\cite{dubey2020understanding}.

\subsection{Cybersickness and curiosity-driven behavior}

Cybersickness and curiosity interact in complex ways to shape user navigation in virtual environments. While cybersickness typically discourages movement by inducing discomfort \cite{shahid2020evaluating,anwar2024moving}, curiosity may drive users to explore novel areas despite potential adverse effects. For instance, users may choose to continue moving forward out of interest in unexplored content, even when early signs of cybersickness appear. Conversely, discomfort may override exploratory impulses, leading to more conservative or avoidance-based behaviors. Understanding how users internally weigh curiosity against discomfort is essential for modeling realistic navigation patterns and optimizing VR system design.

Individual differences further impact this balance. Research has shown that factors such as age, gender, health, and VR experience influence the susceptibility and severity of cybersickness~\cite{davis2014systematic,biswas2024you}. These individual traits also interact with the use of HMDs that worsen discomfort due to sensory conflict between visual and vestibular inputs. Rivera-Flor et al. \cite{rivera2019evaluation} reported that discomfort symptoms arised when using HMD for display affect individual's motivation and attention. Specially, participants reported higher frustration and lower motivation when using the HMD. Familiarity with virtual environments, such as through gaming \cite{dennison2016use}, can mitigate cybersickness by improving spatial prediction and reducing sensory conflict. However, cybersickness remains a critical factor in shaping user behavior and overall satisfaction. Wang et al. \cite{wang2024investigation} found that cybersickness affects walking patterns during navigation, reducing users’ sense of presence. Similarly, Gabel et al. \cite{gabel2023immersive} demonstrated that cybersickness could impair performance in tasks such as text reading, negatively impacting retention and comprehension in educational contexts. Although quantitative models have advanced in predicting cybersickness, they often neglect individual cognitive traits like curiosity \cite{kim2019a,li2023deep}.

In short, users constantly balance between their desire to explore and the risk of discomfort when using VR. This interaction between curiosity and cybersickness is especially critical in self-paced navigation, where neither external incentives nor system constraints dictate movement. Thus, modeling user behavior in VR requires a combination of curiosity-driven motivation and cybersickness-driven aversion.

\section{Methods} \label{sec:methods}

To analyze users' irrational curiosity-driven behavior during virtual navigation, we examined brain decision-making processes in immersive environments using the ReCU model and the iFEP method~\cite{konaka2023decoding}. The ReCU model simulates decision-making by optimizing the expected net utility, while the iFEP method quantitatively decodes this process, estimating key influencing factors. By integrating a curiosity parameter, we accounted for the seemingly irrational decisions made by users, enabling a more nuanced understanding of interaction behavior patterns through quantitative analysis. Furthermore, we investigated how the severity of cybersickness influenced users' navigation strategies, particularly their decisions to accelerate or rest while exploring virtual environments.

\subsection{The ReCU Model}
The ReCU model was designed to replicate the user's decision-making process. It assumes that curiosity and reward are factors that could influence decision-making, where curiosity is one of the reasons a user makes irrational decisions, and reward refers to the credit received after choosing a particular action. In this way, we could discover how VR users determine their navigation speed in an immersive environment, more specifically, balance the desire to explore the VR environment and the severity of cybersickness. The replication process involves two steps. Initially, the VR user adjusts the reward probability for each action based on their observation. Following this, the VR user decides based on the reward probability and curiosity. According to the ReCU model, the VR user seeks to maximize their expected net utility, which is the sum of the expected reward and information gain. The expected net utility  at trial $t$ for choosing action $a_{t+1}$ is determined by
% \begin{equation}
%     U_t(a_{t+1})=E[reward_{t+1}]+c_t \times E[info_{t+1}],
% \end{equation}

\begin{equation}
    U_t(a_{t+1}) = \underbrace{E[reward_{t+1}]}_{
    \substack{\text{recoginition}}\\
    \text{+ reward intensity}
    } + c_t \times \underbrace{E[info_{t+1}]}_{
    \substack{\text{conditional entropy}\\
    \text{+ marginal entropy}}}
\end{equation}

where $c_t$ denotes the curiosity level at trial $t$, $a_t$ denotes the action selected at trial $t$, and $E[\cdot]$ represents the mathematical expectation operator. Curiosity is modeled as a factor that could influence the participant to make irrational decisions and is assumed to fluctuate over time. In our context, reward refers to the presence of cybersickness, which can be seen as a negative reward. In addition, the action selection $a_t$ follows a sigmoid function
\begin{equation}
    P(a_t)=\frac{1}{1+e^{-\beta \Delta U_t}} ,
\end{equation}
where $\beta$ controls the action selection's randomness and $\Delta U_t = U_t(a_{t+1}) - U_t(a'_{t+1})$.

\subsection{The iFEP Method}

The free energy principle (FEP) posits that any system aims to minimize the discrepancy between predictions and actual observations through active inference \cite{friston2010free}. This principle also suggests that systems seek to minimize free energy, which serves as an upper bound for surprise. In the model presented by \cite{konaka2023decoding}, the primary objective is to maximize expected net utility, which corresponds to the negative of expected free energy. A detailed formulation of expected net utility can be found in Appendix \textbf{A}.

To decode the decision-making process of VR users, we employed the inverse Free Energy Principle (iFEP) method, which analyzes behavioral data, including user actions and observations. This decoding process quantifies internal states, such as curiosity levels, recognition of reward probabilities for various options, and confidence in those estimates. The iFEP method operates through a repetitive cycle of predicting internal states followed by corrections, necessitating prolonged trials of behavioral data for the particle filter to converge effectively \cite{konaka2023decoding}.

\subsection{Dataset} \label{sec:dataset}

\begin{figure}
    \centering
    \includegraphics[width=\linewidth]{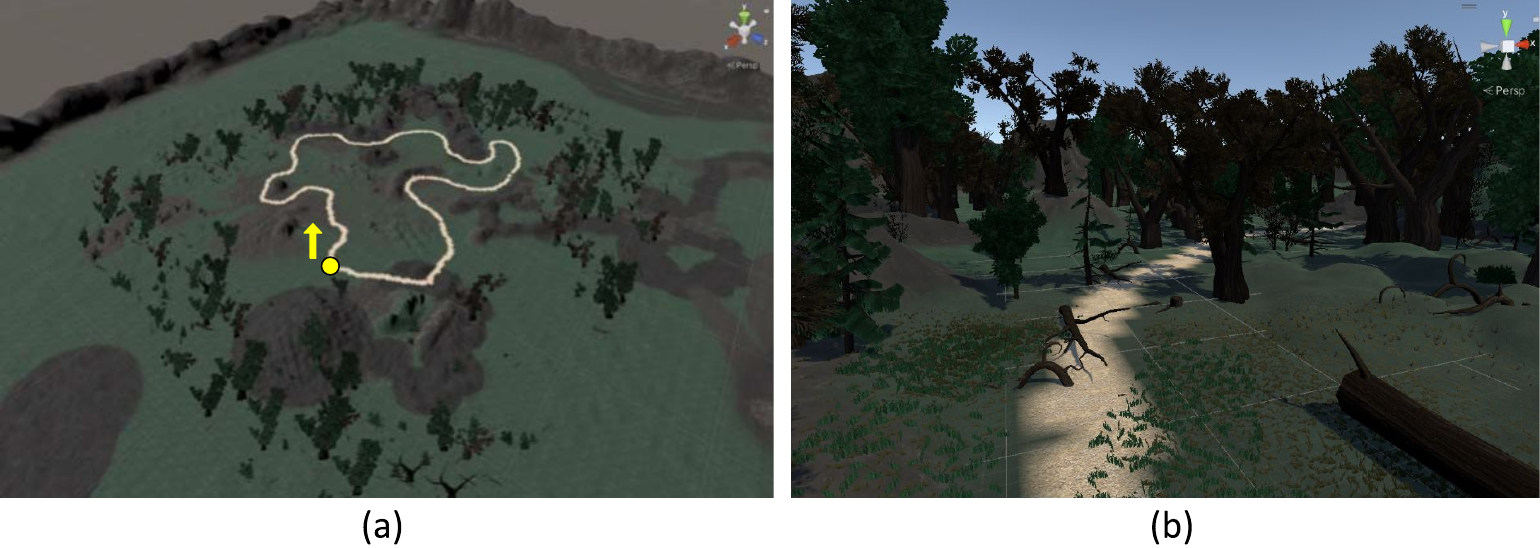}
    \caption{(a) The virtual scenario where the participants perform the navigation task along the highlighted path from the yellow starting point. (b) Detailed environment.}
    \label{fig:navigation_task}
\end{figure}

Sixteen participants (8 females and 8 males; mean age = 25.63 years, SD = 2.91, 25th percentile = 24, 50th percentile = 24.5, 75th percentile = 26) were recruited to perform a navigation task in a VR environment\footnote{\url{https://zenodo.org/doi/10.5281/zenodo.16917139}} (see Figure \ref{fig:navigation_task} and supplmentary video). The scene was designed to focus on individual self-directed exploration. Prior to the experiment, participants completed a health and gaming/VR familiarity survey to report potential issues that could affect the study. The experiment received IRB approval No.HKUST(GZ)-HSP-2024-0003, and informed consent was obtained from all participants.

\begin{figure}
    \centering
    \includegraphics[width=\linewidth]{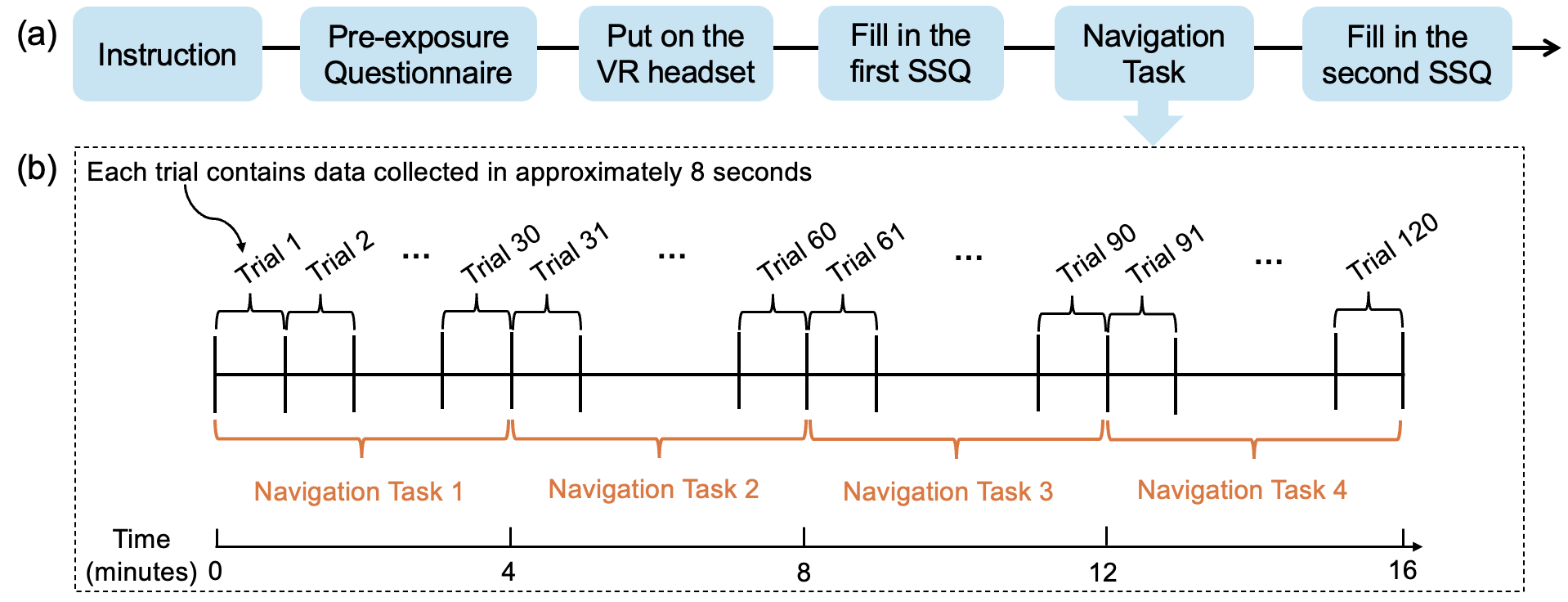}
    \caption{(a) Timeline of the experiment procedure. (b) The detailed procedure of the navigation task. The participants have a short break between each procedure. }
    \label{fig:workflow}
\end{figure}

Each participant completed three navigation tasks on separate days, with a 4- to 5-day interval between each session. Each task lasted 4 minutes, leading to a total duration of 12 minutes. At the end of the experiment, subjects received gifts in recognition of their contribution. Figure \ref{fig:workflow} gives the timeline of the experiment procedure.

At each trial of the navigation task, behavioral data were recorded for modeling with the iFEP method. This included movement speed within the environment, generated by users sliding the touchpad of the VR hand controller, the actions performed, and the severity of cybersickness experienced. Collecting the data enables a more accurate analysis of user behavior.

The experiment procedure is as follows:
\begin{itemize}
    \item The participants received a brief introduction to using the HTC Vive Pro hand controllers to navigate in the virtual environment. Also, they can choose to terminate the experiment early if discomfort occurs.
    \item The participants wear an HTC Vive Pro head-mounted display on their head (to display the virtual environment) and an Empatica E4 wristband on their arm (to collect the electrodermal activity (EDA) signal at a frequency of 4Hz).
    \item The participants enter the virtual environment and navigate using the HTC Vive Pro hand controller touchpad along the highlighted path illustrated in Figure \ref{fig:navigation_task}. To minimize potential confounds in physiological data, participants were instructed to avoid unnecessary speech and body movements during the VR task \cite{marzi2021towards}. Head-tracking (head position and rotation), motion (speed and rotation), and biosignal (EDA, blood volume pulse, temperature, and heart rate) data were collected during the entire process.
    \item The participants completed the SSQ both before and after each navigation task. The difference between post-exposure and pre-exposure contributes to the SSQ score:
    \begin{equation}
        SSQ = SSQ_{post} - SSQ_{pre}
    \end{equation}
    The classification of the severity of cybersickness based on the SSQ score is as follows~\cite{li2023deep}:
        \begin{equation}\label{eq:sick_group}
        SSQ=\left\{\begin{array}{lcl}
        \mathrm{Negligible} &, \mathrm{if} & 0\leq SSQ \leq 5\\
        \mathrm{Low} &, \mathrm{if} & 5< SSQ \leq 20\\
        \mathrm{Moderate} &, \mathrm{if} & 20< SSQ \leq 40\\
        \mathrm{High} &, \mathrm{if} &  SSQ> 40 \\
        \end{array}\right.
        \end{equation}
\end{itemize}

In the navigation task, we categorize the movement of each participant into \textit{Rest} and \textit{Accelerate} based on the average speed every 8 seconds. In addition, we calculate the motion sickness dose value (\textit{MSDV}) \cite{griffin2012handbook} as an estimation of the participant’s cybersickness degree using the following formula:
\begin{equation}
    MSDV = \Bigl(\int_{0}^{T} a^2(t) dt \Bigr)^\frac{1}{2},
\end{equation}
where $T$ is the total duration, and $a$ is the acceleration computed from speed and time. Then, based on the \textit{MSDV} value, it was categorized into \textit{Cybersickness} and \textit{No cybersickness}, which represent feeling cybersickness or not, respectively. Also, the participants wore an Empatica E4 wristband, which could record their EDA data at a rate of 4Hz. The extracted skin conductance response (SCR) signals, which represent the phasic component of the electrodermal activity (EDA) signal, serve as the ground truth for the pwarticipant’s probability of experiencing cybersickness. 

A Chi-square test is conducted to examine the relationship between curiosity and cybersickness based on the participants' SSQ score and curiosity level. The result is statistically significant ($\chi^2=7.27, p=0.026$), implying that the severity of discomfort may influence users’ interaction within the virtual environment. We also conducted a post hoc power analysis based on the observed Chi-square test result to assess the adequacy of our sample size ($N=16, \alpha=0.05$). The calculated effect size (Cram\'er's $V=0.67$) yielded an estimated statistical power of approximately 0.77, indicating a 77\% probability of detecting a true effect. This suggests that the study had sufficient power for the analyses performed.

We performed data preprocessing to fit the iFEP model. First, we utilized the original dataset's speed, EDA signal, and SSQ score. Second, we divided the data collected in one task into 30 intervals. Each interval represents one trial and contains data over 8 seconds. Therefore, each participant's data was segmented into 90 trials, with each 8-second interval chosen to balance the need for robust action classification and alignment with the phasic latency of EDA responses.

The iFEP method takes the participant's action and cybersickness state at each trial as input. It predicts the internal states at each trial, including cybersickness probability, confidence, and curiosity. The internal states model factors that influence a participant to decide whether to go forward or stay still. Therefore, by analyzing these factors, the iFEP method could decode how the participants make decisions by estimating the intensity of curiosity and ground truth reward probability.

\section{Results} \label{sec:results}

This section presents the decoding results after applying the participant’s speed and EDA data to the iFEP method. We conducted a series of tests to evaluate the suitability of using the EDA data in the ReCU model and the iFEP method. In this process, we substituted simulated EDA data into the ReCU model and the iFEP method to analyze the results and validate the approach. After validating the suitability, we substituted the organized user behavioral data into the iFEP method and analyzed the predicted variables.

\subsection{Validation of the ReCU Model}

\begin{figure*}
    \centering
    \includegraphics[width=0.9\linewidth]{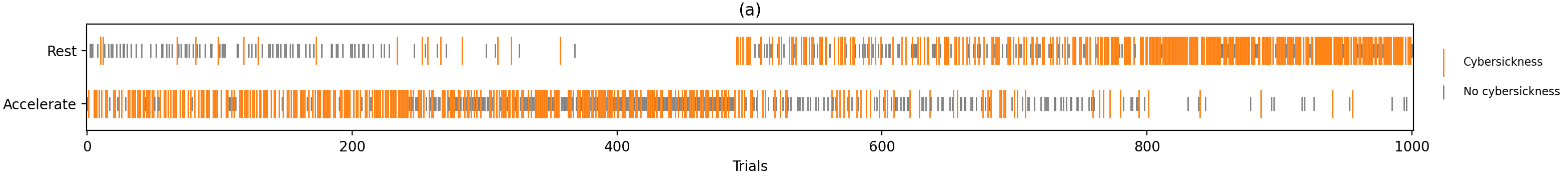}
    \\
    \includegraphics[width=0.8\linewidth]{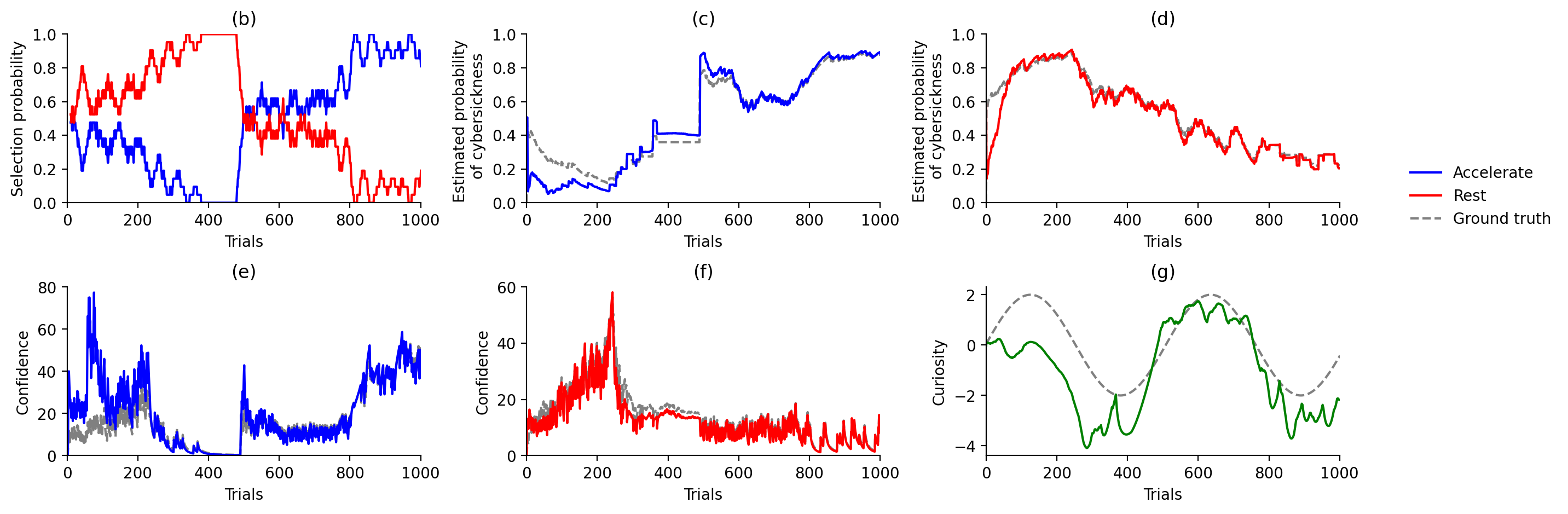}
    \caption{(a) The action selected at each trial. The shorter line indicates no discomfort after selecting an action, whereas the longer line indicates the presence of discomfort after selecting an action. (b) The selection probability for each option. (c, d) The estimated cybersickness probability (solid line) and the ground truth of cybersickness probability (dashed line) for each option. (e, f) The confidence in estimating the probability of cybersickness for each option. (g) The estimated intensity of curiosity (solid line) and the ground truth value of curiosity (dashed line).}
    \label{fig:validate-iFEP}
\end{figure*}

To evaluate the ReCU model's ability to estimate users' cybersickness levels, we generated synthetic EDA signals using the \texttt{eda\_simulate()} function from the \textit{NeuroKit2} library. The simulation test has 1000 trials, and the ground truth of the cybersickness degree is computed as described in Section \ref{sec:methods}. Given the SCR signals extracted from the simulated EDA signals, the ReCU model replicates the decision-making process and estimates the selection probability for each option (\textit{Rest} and \textit{Accelerate}) and the reward probability (i.e., the probability of feeling cybersickness). We assume constant curiosity changes over time and is determined by $c_t=4\sin(4T^{-1}\pi t)$, where $t$ represents the trial number and $T$ represents the total number of trials. While several formulations of $c_t$ were tested and yielded comparable results, we present only this version for clarity. The learning rate $\alpha$ is 0.05, the randomness of action selection $\beta$ is 2, $P_0 = 0.8$, and $\sigma_w=0.4$. Moreover, the ground truth of the cybersickness degree was determined by the average of SCR signals every 8 seconds, and the probability of feeling cybersickness was assumed to be 10\%, 40\%, 60\%, and 90\% to align with the probabilistic outputs of the model. This assigned probability levels are solely used for visualization and interpretive clarity, and do not influence the underlying computational processes.

\begin{figure*}
    \centering\includegraphics[width=0.8\linewidth]{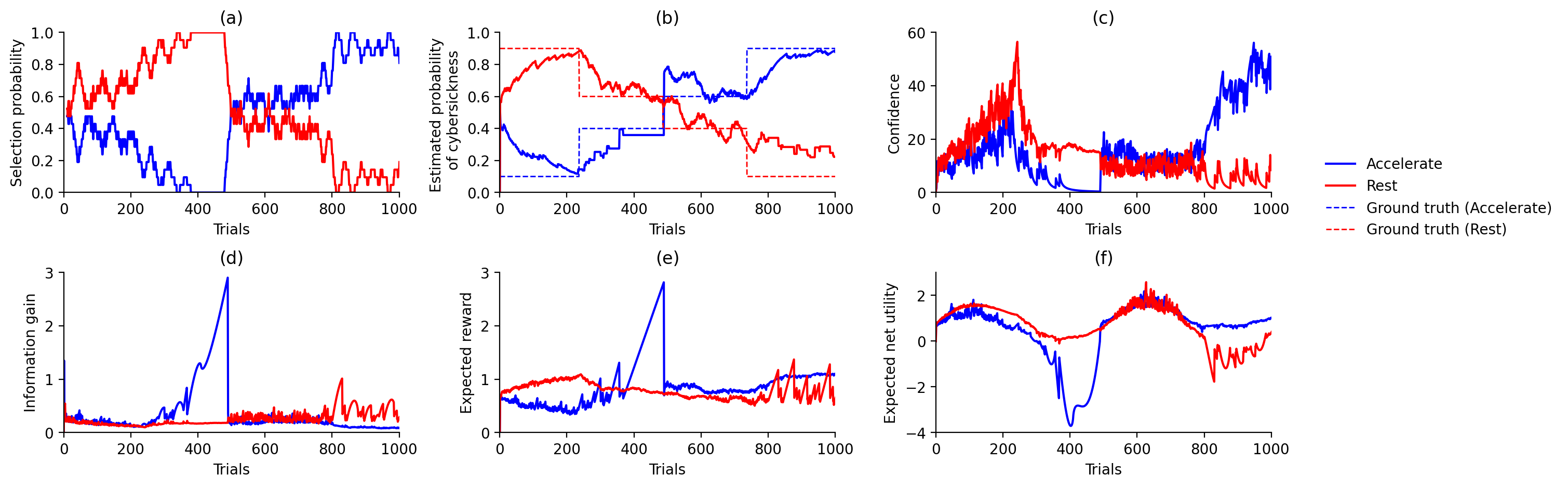}
    \caption{(a) The selection probability for each option. (b) The estimated cybersickness probability for each option, where the solid line indicates the estimated value and the dashed line indicates the ground truth. (c) The confidence in the estimation of the cybersickness probability. (d) The information gained for each option. (e) The expected reward for each option. (f) The expected net utility for each option is the sum of the information gained and the expected reward.}
    \label{fig:validate-ReCU}
\end{figure*}

From Figure \ref{fig:validate-ReCU}b, we can see that the ReCU model can predict the reward probability reasonably well. Also, the confidence in the estimated reward probability for one option increases when that option is selected and vice versa (Figure \ref{fig:validate-ReCU}c). See Appendix \textbf{B} for the formulation of confidence. Additionally, the new information gained may decrease if the same option is selected consecutively (Figure \ref{fig:validate-ReCU}d). Thus, it may encourage the participant to choose another option to obtain more information about the environment. This result shows the validation of the ReCU model for estimating user movement in a VR environment.

\subsection{Validation of the iFEP method}
After that, we used the same simulated EDA signals to test the validity of the iFEP method in the application of decoding the user's decision-making process. First, the extracted SCR signals were input into the ReCU model to estimate the action selection and whether to receive a reward. The estimated reward probability for each option was treated as the ground truth (dashed line in Figures \ref{fig:validate-iFEP}c and \ref{fig:validate-iFEP}d) and compared with the results from the iFEP method. In addition, the estimated action and reward were inputs for the iFEP method to decode the reward probability and intensity of curiosity. The learning rate $\alpha$ is set as 0.05, the inverse temperature $\beta$ is 2, $P_0 = 0.8$ and $\sigma_w = 0.4$. Figure \ref{fig:validate-iFEP}a shows the action selection for 1000 trials, where the shorter line indicates no reward after selecting an action, and the longer line indicates the presence of a reward after selecting an action. Moreover, we can see that the estimated reward probability for each option (red and blue solid lines in Figures \ref{fig:validate-iFEP}c and \ref{fig:validate-iFEP}d) correctly decodes the ground truth (dashed lines in Figures \ref{fig:validate-iFEP}c and \ref{fig:validate-iFEP}d). As a result, Figure \ref{fig:validate-iFEP} shows that the iFEP method could decode user decision-making processes in a VR environment.

\begin{figure*}
    \centering
    \includegraphics[width=0.8\linewidth]{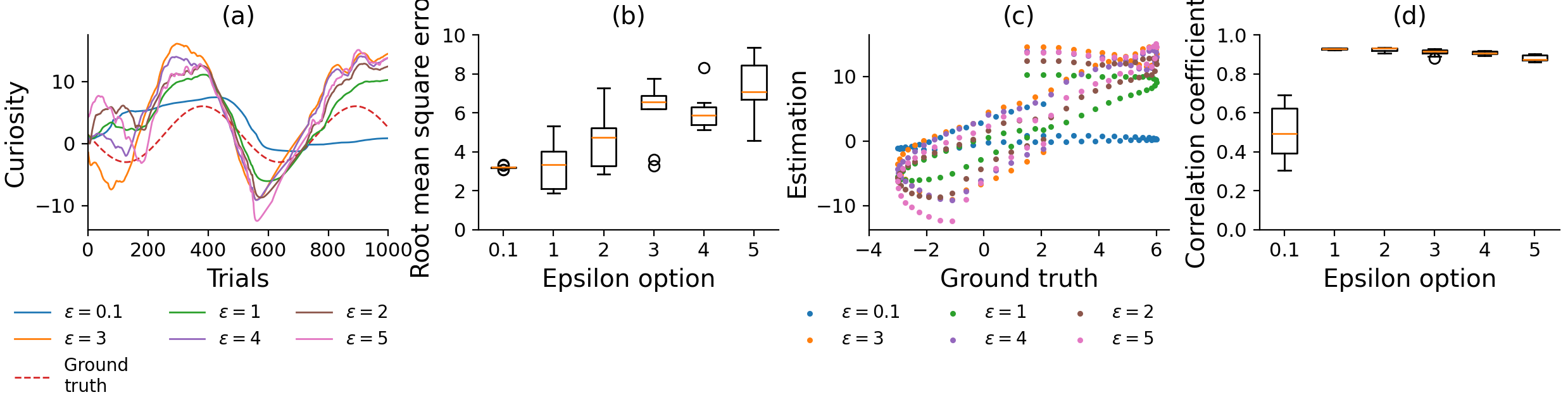}
    \caption{(a) Different predictions of the intensity of curiosity with noise intensity $\varepsilon$ = 0.1, 1, 2, 3, 4, and 5, compared with the ground truth value. (b) The root mean square error with noise intensity $\varepsilon$ = 0.1, 1, 2, 3, 4, and 5 is shown in a box plot. (c) Comparison between the ground truth value and the estimated value of curiosity. (d) A box plot of the correlation coefficient between the estimated and ground truth curiosity with noise intensity $\varepsilon$ = 0.1, 1, 2, 3, 4, and 5.}
    \label{fig:robustness}
\end{figure*}

Furthermore, we investigated the ability of the iFEP method to accurately decode user decision-making processes by applying different noise values to the intensity of curiosity. As shown in Figure \ref{fig:robustness}a, the predictions of curiosity intensity with noise follow a similar pattern to the ground truth. Also, the root mean square error is minor when the value of the noise intensity is smaller (Figure \ref{fig:robustness}b). Moreover, there is a positive correlation between the estimated and the ground truth of curiosity (Figures \ref{fig:robustness}c and \ref{fig:robustness}d). By comparing the estimated and ground-truth curiosity values, we can conclude that the prediction of curiosity intensity is robust under the influence of noise.

\subsection{Results with virtual navigation data}

After validating the ReCU model and iFEP method, we applied real virtual navigation data from the above-mentioned dataset to decode how the user chooses between \textit{Accelerate} and \textit{Rest} resulting from varying curiosity levels. We analyzed how the iFEP method predicted participants' awareness of their likelihood of experiencing cybersickness. Also, we evaluated the participant's navigation behavior in the navigation task. The behavioral data are set as described in Section \ref{sec:dataset}, and the total behavioral data are 90 trials for each participant. Here, we present the results of one participant, whose features are commonly observed in the results of other participants. However, the differences in features in other participants' results will also be taken into account.

\begin{figure*}
    \centering
    \includegraphics[width=\linewidth]{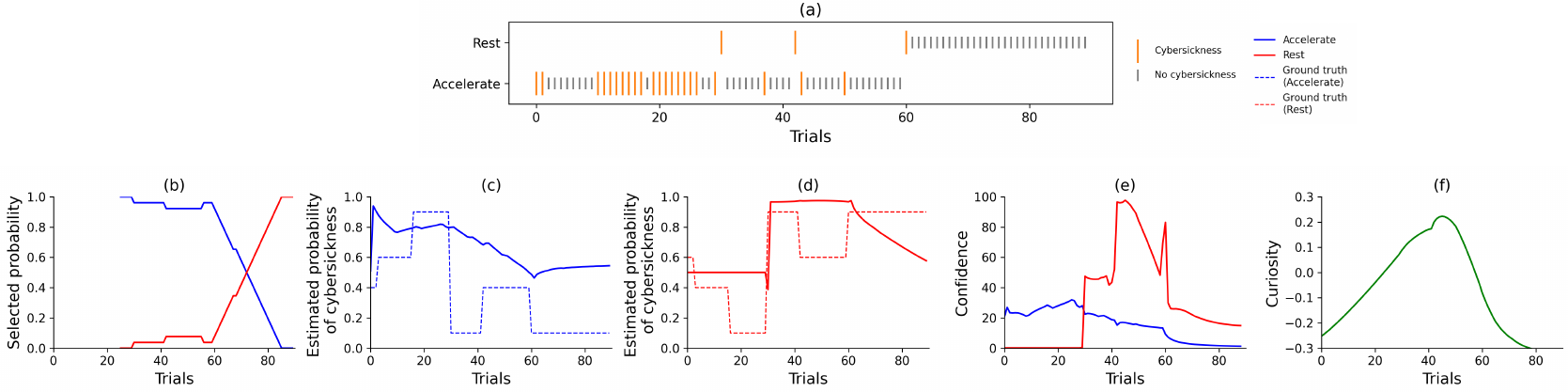}
    \caption{Results from one participant's behavioral data. (a) The presence of cybersickness after selecting an action for 90 trials. The shorter line indicates no discomfort after selecting an action, whereas the longer line indicates the presence of discomfort after selecting an action. Actions with speeds below 1.3m/s are classified as \textit{Rest}, while those above are classified as \textit{Accelerate}. \textit{MSDV} scores below 0.45 are classified as \textit{No Cybersickness}, while those above are classified as \textit{Cybersickness}. (b) The selection probability for each action. (c, d) The estimated cybersickness probability for each action is compared with the ground truth probability. (e) The confidence in the prediction of the cybersickness probability for each action. (f) The prediction of the intensity of curiosity.}
    \label{fig:ifep}
\end{figure*}

Figure \ref{fig:ifep} shows the decoding result of one participant's navigation data using the iFEP method. Figure \ref{fig:ifep}a indicates the action selected and the presence of reward in each trial. Although the estimated values do not perfectly match the ground truth value, the estimations for both actions can recognize the fluctuation in the actual cybersickness probabilities (Figures \ref{fig:ifep}c and \ref{fig:ifep}d). Another reason affecting the accuracy of the estimated value could be that the iFEP assumes a gradual change in reward probability rather than a sudden change \cite{konaka2023decoding}. The confidence in the prediction of the reward probability for choosing to \textit{Rest} and \textit{Accelerate} is demonstrated by Figure \ref{fig:ifep}e. Generally, the intensity of curiosity is negative for most trials, with some positive curiosity around trials 30 to 60, as shown in Figure \ref{fig:ifep}f. According to the SSQ classification, this participant has an average SSQ score of 39.89, indicating a moderate cybersickness. Such an outcome indicates that the participant is relatively conservative in their choice of action during the navigation task. %Despite the general negative curiosity level among most participants, three participants' behavioral data estimates positive curiosity during most trials.
While the majority of participants exhibited overall negative curiosity levels, behavioral data from three individuals consistently demonstrated positive curiosity throughout most experimental trials.

\begin{figure*}
    \centering\includegraphics[width=0.7\linewidth]{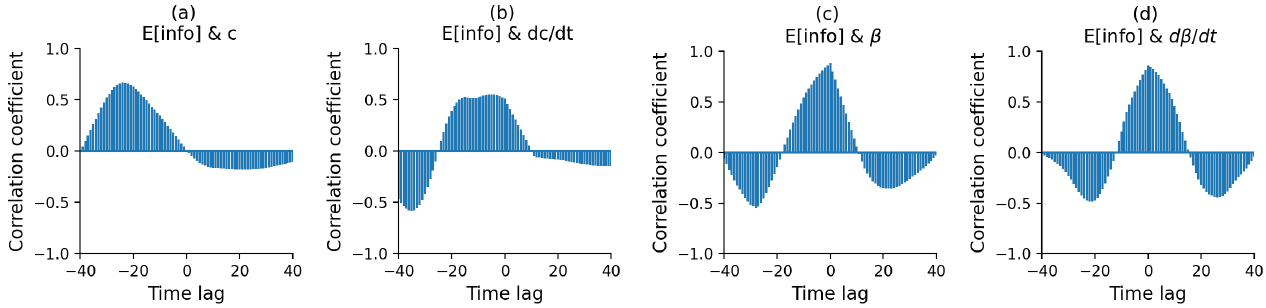}
    \caption{(a) The correlation coefficient between the expected information gain and curiosity at different time lags. (b) The correlation coefficient between the expected information gain and the temporal derivative of curiosity at different time lags. (c) The correlation coefficient between the expected information gain and curiosity at different time lags. (d) The correlation coefficient between the expected information gain and the temporal derivative of curiosity at different time lags.}
    \label{fig:correlation}
\end{figure*}

Additionally, we analyzed user behavior data to investigate the correlation between expected information gain and curiosity. We computed the correlation coefficient between the expected information gain and curiosity and between the expected information gain and the temporal derivative of curiosity. As shown in Figure \ref{fig:correlation}a, the correlation between expected information gain and curiosity at time lag = 0 is minimal (0.0114), indicating no significant relationship between the two variables when they are aligned in time. However, the maximum correlation coefficient, 0.6704, occurs at a time lag of -24 samples. This suggests that, while there is no immediate correlation, a positive correlation emerges when expected information gain lags behind curiosity by 6 seconds. Moreover, the regression line illustrated in Figure \ref{fig:correlation}b shows a positive correlation between the expected information gain and the temporal derivative of curiosity at zero time lag. The maximum correlation for this relationship occurs at a time lag of -35 samples, with a correlation coefficient of -0.5821. When considering the overall dataset from all participants, we observed that the average maximum correlation (accounting for both positive and negative correlations) occurs at a time lag of -2.19 samples, which corresponds to approximately -0.57 seconds. This indicates that, on average, curiosity tends to precede expected information gain by about 0.57 seconds. Thus, it is suggested that curiosity has a predictive role represented by expected information gain. Although the average reaction time for a target detection task is approximately 0.454 seconds \cite{braver2001anterior}, which is faster than the observed lag in our study, this discrepancy can be attributed to individual differences in cognitive processing. Thus, our findings provide a reasonable reaction time, especially in the context of curiosity-related decision-making processes. Therefore, we can conclude that the participant would upscale their curiosity when the expected information gain increases (i.e., the environment becomes unfamiliar).

\subsection{Alternative Methods}
\begin{figure}
    \centering
    \includegraphics[width=0.9\linewidth]{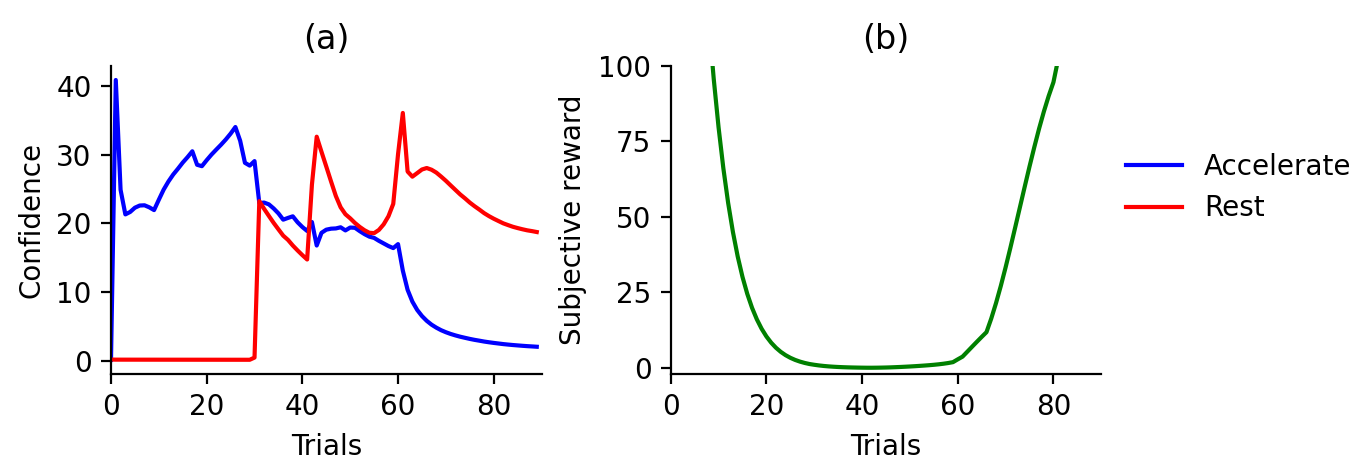}
    \caption{(a) The confidence in the estimated cybersickness probability for each option. (b) The subjective reward at each trial.}
    \label{fig:subjective-reward}
\end{figure}

The subjective reward method and the Q-learning method are two alternative approaches to decoding the user's decision-making process. The subjective reward method states that the expected net utility $U_t$ is computed by
\begin{equation}
    U_t=d_t\times E[reward] + E[info],
\end{equation}
where $d_t$ indicates the desire for reward at trial $t$. Therefore, instead of placing weight on the expected information gain in the iFEP method, the subjective reward method assigns weight to the expected reward. In the context of participants performing navigation tasks in a VR environment, the desire for reward would represent whether the participant would like to experience cybersickness. Therefore, we would expect the value of $d_t$ to be small. However, the average value of reward intensity is 49.08, suggesting that the participant was not motivated to avoid cybersickness (Figure \ref{fig:subjective-reward}b).

On the other hand, the Q-learning method predicts the reward probability $Q$ for action $i$ using
\begin{equation}
    Q_{i,t}=Q_{i,t-1}+\alpha_{t-1} (r_t a_{i,t-1}-Q_{i,t-1}),
\end{equation}
where $\alpha_t$ indicates the learning rate at trial $t$ and $a_{i,t}$ indicates the action selected at trial $t$. Also, the action selection probability is determined by
\begin{equation}
    P(a_{i,t}=1)=\frac{e^{\beta_t Q_{i,t}}}{\sum_{i} e^{\beta_t Q_{i,t}}},
\end{equation}
where $\beta_t$ indicates the inverse temperature at trial $t$ and controls the randomness of action selection. Appendix \textbf{C} gives a detailed explanation of the formulation. Therefore, to understand the relationship between inverse temperature and action selection, we analyzed the correlation coefficient between the expected information gain, inverse temperature, and the derivative of inverse temperature at different time lags.

As shown in Figures \ref{fig:correlation}c and \ref{fig:correlation}d, there is a positive correlation between the expected information gain and inverse temperature (correlation coefficient = 0.8816 at time lag = 0) and between the expected information gain and the temporal derivative of inverse temperature (correlation coefficient = 0.8608 at time lag = 0).

\section{Discussion} \label{sec:discussion}

Our main objective was to model how curiosity and cybersickness influence self-directed navigation in VR. We found that: (1) users adopted predominantly conservative navigation strategies; (2) curiosity increased when the environment changed; (3) iFEP-based estimation captured temporal trends from behavioral data. Most prior research approaches user interaction behavior through psychological or physiological lenses \cite{won2016identifying,gillath2008can,chung2024comparison}. We expand on these perspectives by elucidating the dynamic influence of curiosity in shaping user interactions within immersive environments.

Our results reveal that cybersickness significantly impacts user decisions during navigation. While the model's prediction does not perfectly align with ground truth discomfort, it effectively captures temporal trends in increasing and decreasing probabilities. The gradual nature of these changes may impact prediction precision. Approximately 50\% of participants reported moderate to severe symptoms in SSQ, which is reflected by the reductions in speed. Participants appeared to adjust their strategies to mitigate discomfort, demonstrating self-regulation behaviors. This finding aligns with prior studies on task performance under discomfort~\cite{wang2024investigation, gabel2023immersive}. This suggests that users implicitly manage their health state while navigating, even without explicit instruction to do so. Moreover, participants displayed diverse exploration tendencies, with most people exhibiting conservative behavior likely influenced by cybersickness. However, a subset of three participants, who have a negligible or low SSQ score, showed high curiosity and exploratory actions. These individual differences suggest that curiosity-driven behavior is not solely determined by discomfort, but may also be influenced by other factors worth investigating in future studies.

Explaining the interplay between curiosity and cybersickness in decision-making is complex, leaving multiple interpretations of navigation patterns. First, the secondary task may have distracted participants, exacerbating symptoms \cite{venkatakrishnan2024effects}. Participants often reduced movement speed during navigation, suggesting an implicit health management strategy beyond the primary task. Second, pre-exposure questionnaires indicated that only 31.25\% of participants reported frequent gaming experience. This unfamiliarity with rapidly changing scenes may have contributed to cybersickness, reducing willingness to explore~\cite{tian2022review, wei2019effects}. Additionally, personality traits can shape perception~\cite{ishikawa2021modelling} and responses to novel scenarios~\cite{chen2023eye}, as reflected in the trend toward negative curiosity (Figure \ref{fig:ifep}f) and cautious navigation patterns among most participants. These findings emphasize the importance of examining cybersickness’s influence on individual navigation behaviors, particularly given its prevalence.

% Previous studies associate excessively conservative or curious behaviors with conditions like autism spectrum disorder or attention deficit hyperactivity disorder, where individuals either avoid or seek novel information significantly~\cite{konaka2023decoding}. Our findings reveal a general trend of negative curiosity, with most participants exhibiting cautious navigation behaviors likely influenced by cybersickness. However, three participants deviated, displaying consistently positive curiosity throughout all trials. These individuals, with a negligible and low average SSQ score indicating minimal cybersickness, demonstrated more exploratory behavior. These results suggest that cybersickness is not the sole determinant of navigation patterns, as individuals tolerate discomfort to varying degrees~\cite{davis2014systematic}. In addition, rational behavior reflects a balance between exploratory and reward-seeking actions, where users initially gather information in unfamiliar environments but gradually focus on reward maximization as they gain familiarity~\cite{arnone2011curiosity}. Despite a generally conservative exploratory approach, we observed that users actively sought additional environmental information to reduce uncertainty, aligning with deprivation sensitivity~\cite{noordewier2017curiosity}.

Our findings suggest some possible applications in VR design. First, interpreting users’ internal states allows for systems to adapt their design to respond dynamically to users’ discomfort in real-time, such as adjusting field of view \cite{islam2021cybersickness} and render parameters \cite{uyan2024cdms}. Also, curiosity can serve as an indicator of engagement to stimulate or maintain user engagement in systems designed for learning or training purposes \cite{dubovi2022cognitive}. In addition, user profiling of navigation behavior can be used to set default settings in VR systems \cite{tricomi2023you}. Moreover, our framework may serve as an indicative tool for evaluating VR designs by continuously estimating user comfort and engagement beyond subjective reports.

\subsection{Practicality of alternative models}

The subjective reward method emphasizes the expected reward to model a reward-driven condition. In the context of virtual navigation, this desire for reward reflects participants' inclination to experience or avoid cybersickness. We anticipated that the reward intensity would be low, as participants are likely motivated to avoid discomfort. However, the log average reward intensity across all participants was 2.82. The 25th, 50th, and 75th percentiles were 0.28, 3.77, and 4.71, respectively, indicating a stronger-than-expected desire for reward. Notably, only four participants exhibited a low reward intensity, suggesting they were actively trying to avoid cybersickness. This finding indicates that the subjective reward method may not effectively explain user decision-making in this scenario.

In contrast, the Q-learning method employs reinforcement learning techniques to model adaptive reward-seeking behavior by adjusting time-dependent meta-parameters, focusing exclusively on maximizing expected cumulative rewards. As shown in Figure \ref{fig:correlation}c, the inverse temperature correlates with expected information gain. Additionally, the majority of participants demonstrated a positive correlation between expected cumulative reward and inverse temperature. This result implies that the randomness of action selection aligns with curiosity levels, suggesting that the iFEP method is more suitable for decoding user actions in a VR environment. Regarding the correlation between expected information gain and the temporal derivative of curiosity, results from seven participants showed a positive correlation, two showed a negative correlation, and the remaining seven exhibited no correlation. Given this variability, it is difficult to draw a definitive conclusion about the overall trend.

\subsection{Limitations and future scope}

By investigating the temporal balance between cybersickness and curiosity based on users' virtual navigation behavior, we gain insights into the neural correlates of temporal variability during immersion. However, we are aware that our work may have some limitations. First, the short navigation duration may limit accuracy. The result from the simulated data revealed that longer trials yield more accurate predictions. However, the improved estimation accuracy raises the risk of severe sickness symptoms, complicating the user study~\cite{davis2014systematic}.

Second, the use of a single forest environment allowed control over spatial variables but may have favored certain user preferences, potentially affecting engagement. Future work should explore how environmental context and individual differences (e.g., attention, stress) shape curiosity-driven behavior.

Third, we used a custom VR environment rather than open-source benmarks, as many existing platforms involve passive movement \cite{rouhani2024towards}, multiplayer interaction \cite{fieffer2025mazeworld}, or game-like objectives (e.g., coin collection \cite{milani2023cybersickness}, shooting tasks \cite{calandra2024testbed}) introduce confounding factors. While our setting prioritizes natural navigation and minimal cognitive load, we acknowledge that standardized environments improve comparability.

Finally, other factors (e.g., immersion \cite{martirosov2022cyber}, cognitive distraction \cite{venkatakrishnan2020comparative,jaffe2023modelling}, and repeated exposure \cite{doty2024cybersickness}) may influence behavior but were not explicitly modeled, as we simplified these interactions for modeling purposes in an early stage. Our study included 4-5 day intervals between sessions, which may limit the opportunity for the adaptation effect. Future research should integrate additional physiological and cognitive data, external stimuli, and larger samples to enhance robustness and generalizability.

\section{Conclusion}

This research presents a computational framework for understanding how curiosity and cybersickness jointly influence user behavior during virtual navigation. We find that most users tend to adopt cautious exploratory strategies when experiencing discomfort. The iFEP modeling reveals a strong link between expected information gain and curiosity, demonstrating how users manage uncertainty and discomfort simultaneously.

Our findings highlight the importance of designing VR systems that could respond to individual performance and engagement. By modeling internal variables, we move closer to user-centered design that enhances user experience, especially in personalized learning, training, and entertainment within immersive VR environments.

As immersive technologies advance, understanding the tradeoffs between cognitive drives and sensory experience is crucial. Future work can incorporate real-time adaptation strategies that adjust interaction dynamics based on user state estimation to provide timely feedback and enhance user experience.

\section*{Acknowledgments}
This work was supported by the Guangzhou-HKUST(GZ) Joint Funding Scheme (2025A03J3955) and the Science and Technology Planning Project of Guangdong Province (2024GXJK10).

{\appendices
\section{Expected net utility} \label{appendix:expected net utility}
This section provides the calculation process for the expected net utility according to \cite{konaka2023decoding} and \cite{parr2019generalised}.
The expected net utility is calculated using
\begin{equation} \label{eq:expected net utility}
    U_t(a_{t+1})=E[reward_{t+1}] + c_t \times E[info_{t+1}],
\end{equation}
where $a_t$ is action and $c_t$ is curiosity intensity at trial $t$. In addition, the cybersickness probability is influenced by the latent cause $w$ and the participant’s action $a$. Moreover, the latent cause
\begin{equation}
    w_{i,t} = w_{i,t-1} + \sigma_{w} \times \epsilon_{i,t}, 
\end{equation}
where $\sigma_{w}$ is the noise intensity, $\epsilon_{i,t}$ is the standard Gaussian noise, and $i$ represents the option's index ($i=1$ represents the action \textit{Rest}, and $i=2$ represents the action \textit{Accelerate}). Therefore, the cybersickness probability is computed using
\begin{equation}
    f(w_{i,t}) =  \frac{1}{1+e^{-w_{i,t}}}.
\end{equation}
The participant’s recognition of the environment is represented as the following equation \cite{konaka2023decoding}:
\begin{equation} \label{eq:recognition}
    P(o_t|\mathbf{w}_t,\mathbf{a}_t) = \prod_{i} \Bigr[ f(w_{i,t})^{o_t} (1-f(w_{i,t}))^{1-o_t} \Bigl] ^{a_{i,t}},
\end{equation}
where $o_t\in \{0,1\}$ represents whether cybersickness occurs given the latent variable $\mathbf{w}_t = (w_{1,t},w_{2,t})^T$ and action $\mathbf{a}_{t} \in \bigl( (1,0)^T, (0,1)^T \bigr)$ at trial $t$.
The Taylor series expansion for a real and differentiable function $f(x)$ at the point $x=a$ is a linear approximation at $x=a$, where
\begin{equation}
    f(x)=\sum_{n=0}^{\infty} \frac{f^{(n)}(a)}{n!}(x-a)^n.
\end{equation}
Therefore, we obtain the following equation by applying the 2nd-order Taylor series expansion on equation (\ref{eq:recognition}):
\begin{gather}
    P(\mathbf{o}_{t+1}|\mathbf{a}_{t+1})=\int P(o_{t+1}|\mathbf{w}_{t+1},\mathbf{a}_{t+1})Q(\mathbf{w}_{t+1}|\mathbf{a}_{t+1}) d\mathbf{w}_{t+1} \nonumber \\
    =\int \prod_{i}\Bigl( f(w_{i,t+1})^{o_{t+1}}(1-f(w_{i,t+1})^{1-o_{t+1}}) \Bigr)^{a_{i,t+1}} \nonumber \\
    Q(\mathbf{w}_{t+1}|\mathbf{a}_{t+1})d\mathbf{w}_{t+1} \nonumber \\
    =\prod_{i} \Bigl( f(w_{i,t+1})^{o_{t+1}}(1-f(w_{i,t+1})^{1-o_{t+1}}) \nonumber \\
    +1^{o_{t+1}}(-1)^{1-o_{t+1}}\frac{1}{2}f(\mu_{i,t+1})(1-f(\mu_{i,t+1})) \nonumber \\
    (1-2f(\mu_{i,t+1}))(p_{i,t}^{-1}+p_{w}^{-1}) \Bigr) ^{a_{i,t+1}}. \label{eq:POA}
\end{gather}
$Q(\mathbf{w}_t|\varphi_t) = N(\mathbf{w}_t|\boldsymbol\mu_{t},\boldsymbol\Lambda_{t}^{-1})$ is a Gaussian distribution where $\varphi_{t} = (\boldsymbol\mu_{t}, \boldsymbol\Lambda_{t})$. In addition, $\boldsymbol\mu_{t}=(\mu_{1,t},\mu_{2,t})^T$ represents the mean and $\boldsymbol\Lambda_{t}=diag(p_{1,t},p_{2,t})$ represents the precision.  Based on the participant’s desired probability of cybersickness occurrence $P_0$, the reward intensity is
\begin{equation}
    R = \begin{cases} 0, & \text{if $o_t = 0$, i.e. no cybersickness occurs.} \\
    \ln \frac{P_0}{1-P_0}, & \text{if $o_t = 1$, i.e. cybersickness occurs.}\end{cases}
\end{equation}
Then, the expected reward (the first term of equation (\ref{eq:expected net utility})) can be formulated as follows:
\begin{align} \label{eq:expected reward}
    E[reward_{t+1}] &= E_{P(o_{t+1}|\mathbf{a}_{t+1})}[R(o_{t+1})] \nonumber \\
    &= \sum_{i} P(o_{t+1}|\mathbf{a}_{i,t+1})R(o_{t+1}).
\end{align}
The expected information gain (the second term of equation (\ref{eq:expected net utility})) is computed using
\begin{align} \label{eq:expected info}
    E&[info_{t+1}] \nonumber \\
    &=E_{P(o_{t+1}|\mathbf{a}_{t+1})} \Bigl[ D_{KL}[Q(\mathbf{w}_{t+1}|o_{t+1}, \mathbf{a}_{t+1})||Q(\mathbf{w}_{t+1}|\mathbf{a}_{t+1})] \Bigr] \nonumber \\
    &= H(o_{t+1}) - H(o_{t+1}|\mathbf{w}_{t+1}),
\end{align}
where Kullback-Leibler divergence ($D_{KL}$) measures the difference between probability distrbutions $Q(\mathbf{w}_{t+1}|o_{t+1},\mathbf{a}_{t+1})$ and $Q(\mathbf{w}_{t+1}|\mathbf{a}_{t+1})$, also known as information gain \cite{parr2019generalised}. The first term $H(o_{t+1})$ is the marginal entropy and the second term $H(o_{t+1}|\mathbf{w}_{t+1})$ is the conditional entropy. First, the marginal entropy is
\begin{gather}
    H(o_{t+1}) = E_{P(o_{t+1}|\mathbf{a}_{t+1})}[-\ln P(o_{t+1}|\mathbf{a}_{t+1})] \nonumber \\
    = -\sum_{i} a_{i,t+1}
    \Bigl(P(o_{t+1}=0|\mathbf{a}_{t+1}) \ln P(o_{t+1}=0|\mathbf{a}_{t+1}) \nonumber \\
    + P(o_{t+1}=1|\mathbf{a}_{t+1}) \ln P(o_{t+1}=1|\mathbf{a}_{t+1}) \Bigr). \label{eq:marginal_ent1}
\end{gather}
We substitute equation (\ref{eq:POA}) into equation (\ref{eq:marginal_ent1}) to solve the formula for the marginal entropy.
The conditional entropy is
\begin{gather}
    H(o_{t+1}|\mathbf{w}_{t+1}) = \nonumber \\
    E_{P(o_{t+1}|\mathbf{w}_{t+1}, \mathbf{a}_{t+1})Q(\mathbf{w}_{t+1}|\mathbf{a}_{t+1})}) [\ln P(o_{t+1}|\mathbf{w}_{t+1},\mathbf{a}_{t+1})] \nonumber \\
    =-E_{Q(\mathbf{w}_{t+1}|\mathbf{a}_{t+1})} \Bigl[ \ln \prod_{i} \Bigl( f(w_{i,t+1})^{o_{t+1}}\nonumber  \\
    (1-f(w_{i,t+1})^{1-o_{t+1}})^{a_{i,t+1}} \Bigr) \Bigr] \nonumber \\
    =-E_{Q(\mathbf{w}_{t+1}|\mathbf{a}_{t+1})} \Bigl[ \sum_{i} a_{i,t+1} \Bigl( f(w_{i,t+1}) \ln f(w_{i,t+1}) \nonumber \\
    + (1-f(w_{i,t+1})) \ln (1-f(w_{i,t+1})) \Bigr) \Bigr]
\end{gather}
We obtain the above equation by substituting equation (\ref{eq:recognition}) to $H(0_{t+1}|\mathbf{w}_{t+1})$. Then, to approximate the value of conditional entropy, we would use Taylor’s theorem. Let $g(w_{i,t+1}) = f(w_{i,t+1}) \ln f(w_{i,t+1}) + (1-f(w_{i,t+1})) \ln (1-f(w_{i,t+1})$ and apply the 2nd-order Taylor series expansion to it. Therefore, we obtain
\begin{gather}
    H(o_{t+1}|\mathbf{w}_{t+1}) \approx - \sum_{i}a_{i,t+1} \Bigl[ f(w_{i,t+1}) \ln f(w_{i,t+1}) \nonumber \\
    + (1-f(w_{i,t+1})) \ln (1-f(w_{i,t+1})) \nonumber \\
    + \frac{1}{2} \Bigl(f(\mu_{i,t+1})(1-f(\mu_{i,t+1}))(1+(1-2f(\mu_{i,t+1})) \nonumber \\
    \ln \frac{f(\mu_{i,t+1})}{1-f(\mu_{i,t+1})} )\Bigr) (p_{i,t}^{-1}+p_w^{-1}) \Bigr]. \label{eq:conditional ent}
\end{gather}
Afterward, we need to compute the value of curiosity. Assuming that the curiosity varies at each trial and is influenced by curiosity's noise $\zeta$, the formula of curiosity at trial $t$ is:
\begin{equation} \label{eq:curiosity}
    c_{t} = c_{t-1} + \epsilon_{c} \times \zeta_{t},
\end{equation}
where $\epsilon_{c}$ is the noise intenisty \cite{konaka2023decoding}.
Therefore, by substituting equations (\ref{eq:marginal_ent1}) and (\ref{eq:conditional ent}) into equation (\ref{eq:expected info}) we obtain the value of the expected information gain. Finally, we attain the value of the expected net utility by substituting the expected reward (equation (\ref{eq:expected reward})), expected information gain (equation (\ref{eq:expected info})) and curiosity intensity (equation (\ref{eq:curiosity})) to equation (\ref{eq:expected net utility}).

\section{Confidence} \label{appendix:confidence}
The confidence of cybersickness probability recognition is
\begin{equation}
    \gamma_{i,t} = \frac{p_{i,t}}{f'(\mu_{i,t})^2},
\end{equation}
where $f$ is the cybersickness probability and $p_{i,t}$ is the precision of the probability distribution Q. Moreover, the value of $p_{i,t}$ is updated according to:
\begin{equation}
    p_{i,t} = K_{i,t}^{-1}+f(\mu_{i,t})(1-f(\mu_{i,t})),
\end{equation}
and
\begin{equation}
    K_{i,t} = \sigma_w^{2} + p_{i,t}^{-1}.
\end{equation}

\section{The Q-learning Method} \label{appendix:qlearning}
The Q-learning method uses the following action value function to predict the reward obtained at trial $t$ and choosing action $i$:
\begin{equation}
    Q_{i,t} = Q_{i,t-1} + \alpha_{t-1}(r_{t}a_{i,t-1} - Q_{i,t-1}),
\end{equation}
where $\alpha_t$ represents the learning rate. Additionally, the softmax function 
\begin{equation}
    P(a_{i,t}=1) = \frac{e^{B_{t}Q_{i,t}}}{\sum_{i}e^{B_{t}Q_{i,t}}}, 
\end{equation}
where $B_t$  represents the inverse temperature and controls the randomness of action selection, is used to determine the action selected by the participant. The learning rate and inverse temperature values are calculated using behavioral data at each time point. These parameters are assumed to change temporally and follow
\begin{equation}
    \theta_{t} = \theta_{t-1} + \epsilon_{\theta}\times\zeta_{\theta,t},
\end{equation}
where $\theta \in \{ \alpha,\beta \}$, $\epsilon_{\theta}$ represents the noise intenisty, and $\zeta_{\theta,t}$ represents the white noise. This allows for adjustments based on the participant's actions and outcomes. Consequently, the reward prediction at a given trial depends on the learning rate, inverse temperature, and the predicted reward from the previous trial.
}

% \bibliographystyle{IEEEtran}
% \bibliography{IEEE_THMS_Accepted/Reference}
%


\vfill

\end{document}